\documentclass[pra,showpacs,twocolumn,superscriptaddress,floatfix,aps]{revtex4-2}
\usepackage{amsmath,amssymb,bm}
\usepackage{graphicx}
\usepackage{physics}
\usepackage{color}
\usepackage{soul}
\usepackage{mathtools}
\usepackage[utf8]{inputenc}
\usepackage[T1]{fontenc}
\usepackage{array}
\usepackage{float}
\usepackage{comment}
\usepackage{appendix}
\begin{document}
\title{Cavity optomechanical detection of persistent currents and solitons in a bosonic ring condensate}
\author{Nalinikanta Pradhan}
\affiliation{Department of Physics, Indian Institute of Technology, Guwahati 781039, Assam, India}
\author{Pardeep Kumar}
\affiliation{Max Planck Institute for the Science of Light, Staudtstra\ss e 2, 91058 Erlangen, Germany}
\author{Rina Kanamoto}
\affiliation{Department of Physics, Meiji University, Kawasaki, Kanagawa 214-8571, Japan}
\author{Tarak Nath Dey}
\affiliation{Department of Physics, Indian Institute of Technology, Guwahati 781039, Assam, India} 
\author{M. Bhattacharya}
\affiliation{School of Physics and Astronomy, Rochester Institute of Technology, 84 Lomb Memorial Drive, Rochester, New York 14623, USA}
\author{Pankaj Kumar Mishra}
\affiliation{Department of Physics, Indian Institute of Technology, Guwahati 781039, Assam, India}

\date{\today}

\begin{abstract} 
We present numerical simulations of the cavity optomechanical detection of persistent currents and bright solitons in an atomic Bose-Einstein condensate confined in a ring trap. This work describes a novel technique that measures condensate rotation \textit{in situ}, in real-time, and with minimal destruction, in contrast to currently used methods, all of which destroy the condensate completely. For weakly repulsive inter-atomic interactions, the analysis of persistent currents extends our previous few-mode treatment of the condensate [P. Kumar \textit{et al}. Phys. Rev. Lett. \textbf{127}, 113601 (2021)] to a stochastic Gross-Pitaevskii simulation. For weakly attractive atomic interactions, we present the first analysis of optomechanical detection of matter-wave soliton motion. We provide optical cavity transmission spectra containing signatures of the condensate rotation, sensitivity as a function of the system response frequency, and atomic density profiles quantifying the effect of the measurement backaction on the condensate. We treat the atoms at a mean-field level and the optical field classically, account for damping and noise in both degrees of freedom, and investigate the linear as well as nonlinear response of the configuration. Our results are consequential for the characterization of rotating matter waves in studies of atomtronics, superfluid hydrodynamics, and matter-wave soliton interferometry.

\end{abstract}

\flushbottom

\maketitle
\section{Introduction}
An atomic Bose-Einstein Condensate (BEC) confined in a ring potential exhibits superflow, {\it i.e.}, transport without dissipation \cite{RyuPRL2007,GuoPRL2020,GuoNJP2022}. It is therefore a natural platform for studying superfluid hydro-dynamical phenomena such as quantized persistent currents \cite{MoulderPRA2012}, phase slips \cite{WrightPRL2013,SnizhkoPRA2016}, excitations \cite{WrightPRA2013}, two-component rotation \cite{BeattiePRL2013,GallemiNJP2015}, hysteresis \cite{EckelNature2014,HouPRA2017}, and shock waves \cite{WangNJP2015}; a versatile enabler for applications such as matter-wave interferometry \cite{GuptaPRL2005,MartiPRA2015}, atomtronic circuits \cite{RamanathanPRL2011,RyuPRL2013,PandeyPRL2021,AmicoRMP2022}, and gyroscopy \cite{CooperPRA2010,PelegriNJP2018}; and a convenient simulator of topological excitations \cite{KanamotoPRL2008,DasScientificReports2012,CormanPRL2014}, early universe cosmology \cite{EckelPRX2018}, and time crystals \cite{OhbergPRL2019}. 

Inspired by the experimental activity in the field, a large number of theoretical proposals have been put forward, based on the BEC-in-a-ring system, characterizing plain wave to soliton transitions \cite{KanamotoPRA2003}, self-trapping \cite{BaharianPRA2013}, simulated Hawking radiation \cite{YatsutaPRR2020}, the Berry phase \cite{TodoricPRA2020}, qubits for computation \cite{AghmalyanNJP2015}, critical velocities \cite{PiazzaJPB2013,ArabahmadiPRA2021}, superflow decay \cite{MatheyPRA2014, KunimiPRA2019,PoloPRL2019,MehdiSPP2021}, phonons \cite{ModugnoPRA2006}, rotating lattices \cite{HuangPRA2021}, rotation sensing \cite{RagolePRL2016}, gauge fields \cite{CominottiPRL2014}, matter-wave interference \cite{KialkaPRR2020}, double-ring geometries \cite{BrandPRA2009,OliinykJPB2019,ObiolPRR2022}, etc. 

In all these studies, knowledge of the condensate rotation is an important consideration. At present, all demonstrated methods of detecting such rotation in ring BECs are destructive of the condensate \cite{KumarNJP2016}. Due to issues related to optical resolution, the methods typically also require time-of-flight expansion of the atoms, making \textit{in situ} measurements difficult. A theoretical proposal exists based on atom counting for a minimally destructive measurement of the condensate rotation \cite{SafaeiPRA2019}.

Recently, our group suggested a method for detecting condensate rotation in real time, \textit{in situ} and with minimal destruction to the condensate \cite{KumarPRL2021}. This method proposed to use the techniques of cavity optomechanics, a discipline that addresses the coupling of mechanical motion to electromagnetic fields confined in resonators \cite{AspelmeyerRMP2014}. Probably the best-known optomechanical device in existence is the Laser Interferometer Gravitational-Wave Observatory (LIGO), which detected the gravitational waves predicted by Einstein's theory of general relativity \cite{AbbottPRL2016}, an accomplishment recognized by a Nobel prize. 

Cavity optomechanics is now a mature field that is capable of supporting the sensitive detection of any physical variable that actuates the mechanical motion coupling to the electromagnetic fields in the cavity. Thus, cavity optomechanical principles have been employed to construct accelerometers \cite{KrauseNP2012}, magnetometers \cite{ForstnerPRL2012}, thermometers \cite{MontenegroPRR2020}, mass \cite{SansaNC2020} and force \cite{FoglianoNC2021} sensors, etc. In our previous proposal, which considered a rotating BEC in a cavity, it was shown that the resulting sensitivity of BEC rotation measurement was three orders of magnitude better than demonstrated hitherto \cite{KumarPRL2021}. This conclusion regarding the detection of a persistent current was based on a few-mode approximation for the condensate.

In the present work, we consider a BEC confined in a ring trap and interacting with an optical cavity mode carrying orbital angular momentum (OAM) \cite{KumarPRL2021}. This may be regarded as the rotational analog of a BEC with a linear degree of mechanical freedom combined with a standing wave optical cavity lattice in an optomechanical context \cite{BrenneckeScience2008}. 

For weak repulsive atomic interactions, we extend the previous two-mode characterization of the condensate to a mean field, {\it i.e.}, Gross-Pitaevskii, treatment. Our method allows us to confirm the basic results of the two-mode treatment regarding the rotation detection of a persistent current, to investigate the modifications resulting from taking the full condensate dynamics into account, and to quantify the effect of measurement backaction on the condensate. It also allows us to consider the detection of a superposition of persistent current states in the condensate.

We also investigate, for the first time, the case of weak attractive atomic interactions \cite{KavoulakisPRA2003} - which results in a bright-soliton ground state in the ring condensate - in the optomechanical context. Such solitons are of great interest e.g., to rotation sensing and matter-wave interferometry \cite{BrandJPB2001,ToikkaJPB2013,McDonaldPRL2014,HelmPRL2015,GalluciNJP2016,
JezekPRA2016,CataldoEPJD2016}. However, a soliton is not amenable to a few-mode optomechanical treatment, due to the large number of matter-wave OAM states contributing to the condensate dynamics. Our numerical simulations make this case tractable, extracting, as in the case of the persistent currents, cavity transmission spectra with signatures of soliton rotation, the sensitivity of the measurement as a function of system response frequency, and atomic density profiles showing the effect of the measurement on the condensate. In all simulations, the matter is treated at the mean-field level, light is treated classically, and noise arising from both optical as well as matter-wave fields are taken into account.  

This paper is organized as follows. In Section~\ref{TheorModelNumSim} the theoretical model and details of the numerical simulation are presented. In Sections \ref{PersistentCurrent} and \ref{Soliton} we provide the dynamics, OAM content, optical spectra, measurement sensitivity, and condensate density fidelity for the persistent current and bright soliton detection, respectively. The conclusions are presented in Section \ref{sec:5}.


\section{Theoretical model and details of numerical simulation}
\label{TheorModelNumSim}
\begin{figure}[!htp]
\begin{center}
    \includegraphics[width=1.0\linewidth]{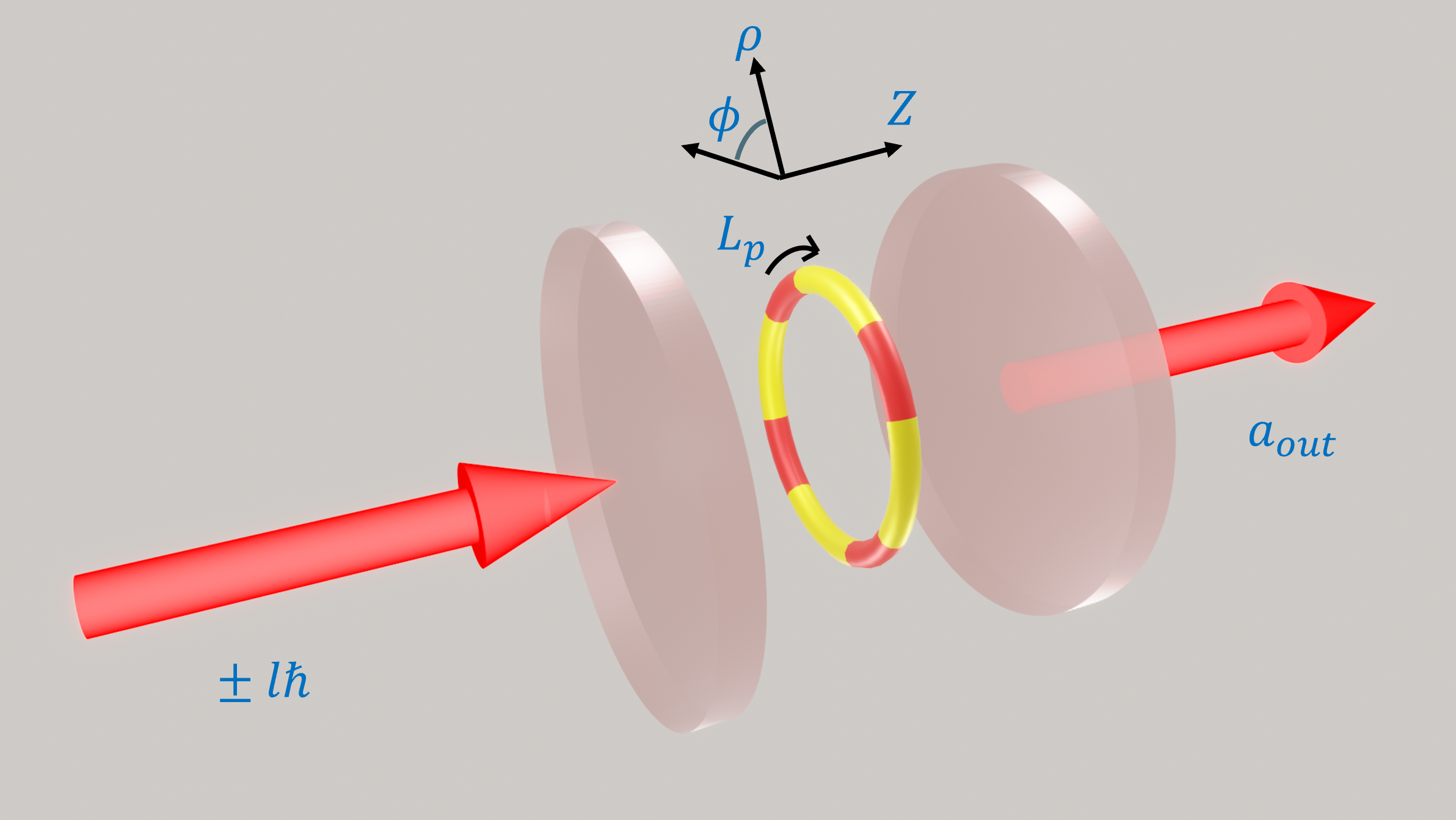} \\

\end{center}
\caption{A schematic setup for the BEC with winding number $L_p$ rotating in a ring trap around the axis of the Fabry-Perot cavity. The red beam represents the Laguerre-Gauss modes with the orbital angular momentum of $\pm l \hbar$ used to probe the BEC rotation. The output signal $a_{out}$ is the field transmitted from the cavity.}
\label{fig:setup}
\end{figure}

In this section, we present the theoretical model for the configuration of interest, shown in Fig.~\ref{fig:setup}, {\it i.e.}, a BEC confined in a one-dimensional ring trap coupled to a cavity using Laguerre-Gauss beams \cite{KumarPRL2021}. 

The dynamical equations governing the system are given by \cite{blakie2008dynamics, stoof2001dynamics, brennecke2008cavity,KumarPRL2021}
\begin{equation}
\begin{aligned}
    (i-\Gamma)\frac{d\psi}{d\tau} = \biggl[-  \frac{d^2}{d\phi^2}& + \frac{U_0}{\omega_\beta}  |\alpha(\tau)|^2 \cos^2\left({l\phi}\right) \\ & -\mu   + 2 \pi \frac{\chi}{N}|\psi|^2\biggr] \psi + \xi(\phi,\tau),
    \label{Eq:bec}    
\end{aligned}
\end{equation}
and
\begin{equation}
\begin{aligned}
    i\frac{d\alpha}{d\tau} = \biggl\{ - \biggl[\Delta_c &- U_0 \langle \cos^{2}\left(l\phi\right)\rangle_{\tau} + i \frac{\gamma_{0}}{2}\biggr] \alpha \\ & + i\eta \biggr\} \omega_\beta^{-1} + i \sqrt{\gamma_0} \omega_\beta^{-1} \alpha_{in}(\tau). 
    \label{Eq:cavity}
\end{aligned}
\end{equation}

In the above, Eq.~(\ref{Eq:bec}) is the stochastic Gross-Pitaevskii equation, where $\psi \equiv \psi(\phi, \tau)$ represents the microscopic wave function of the condensate with $\phi$ the angular variable along the ring, and $\tau$ the scaled time, to be
defined below. The wave function obeys, at any time, the normalization condition
\begin{equation*}
    \int_{0}^{2 \pi} |\psi(\phi,\tau)|^2  d\phi = N,
\end{equation*}
where $N$ is the number of atoms in the condensate. In order to obtain the dimensionless Eq.~(\ref{Eq:bec}), energy and time have been scaled using the quantities 
\begin{equation}
    \hbar \omega_{\beta} = \frac{\hbar^2}{2 m R ^2} \,\, \mathrm{and}
    \,\,\tau = \omega_\beta t,
\end{equation}
respectively, where $m$ is the atomic mass and $R$ is the radius of the ring-shaped trap. 

The first term inside the square bracket on the right-hand side of Eq.~(\ref{Eq:bec}) stands for the kinetic energy of the atoms due to their rotational motion. The second term in the bracket represents the optical lattice potential created with a superposition of two Laguerre-Gauss beams having orbital angular momenta $\pm l\hbar$ respectively, with $U_0 = g_0^2 / \Delta_a$, where $g_{0}$ is the single photon-single atom coupling and $\Delta_{a}$ is the detuning of the driving laser from the atomic resonance. The third term in the bracket corresponds to the chemical potential $\mu$ of the condensate, which is corrected by $\Delta \mu$ at each time step ($\Delta \tau$) as~\cite{mithun2018signatures}
\begin{equation*}
    \Delta\mu = (\Delta \tau)^{-1} \ln{\left[\int \lvert \psi(\phi, \tau)\rvert^2 d\phi / \int \vert\psi(\phi, \tau + \Delta \tau)\rvert^2 d\phi\right]},     
\end{equation*}
to conserve the normalization of the condensate in the presence of the dissipation
\begin{equation}
\Gamma=\frac{\omega_{m}}{\omega_{\beta}},    
\end{equation}
set by the lifetime $\omega_{m}^{-1}$ of the persistent currents \cite{RyuPRL2007,KumarPRL2021}. 
The fourth term inside the bracket represents the scaled atomic interaction
\begin{equation}
\chi  = \frac{gN}{2 \pi \hbar \omega_\beta}.    
\end{equation}
Here 
\begin{equation}
g = \frac{2\hbar \omega_\rho a_s}{R},
\end{equation}
with $a_s$ the $s$-wave atomic scattering length and $\omega_\rho$ the harmonic trap frequency along the radial direction \cite{KumarPRL2021}. Thermal noise $\xi$, with zero mean and correlations provided below, has been added to the condensate in accordance with fluctuation-dissipation theory \cite{kubo1966fluctuation}. 

The dynamics of the complex intracavity coherent field amplitude $\alpha$ is described by Eq.~(\ref{Eq:cavity}). In our simulations, we have treated $\alpha$ as a classical quantity as this approximation has been shown to be adequate for similar setups experimentally. For example, in \cite{BrenneckeScience2008} although bistability is observed at intracavity photon numbers below unity $(|\alpha|^{2}\lesssim 1)$, the corresponding experimental data is very well described using a classical theory for the optical field \cite{RitterAPB2009}. As explained in \cite{Brenneckethesis}, this is due to the fact that in the `bad cavity' limit \cite{AspelmeyerRMP2014}, where the condensate mechanical (i.e., sidemode) oscillation frequencies are lower than the cavity linewidth, the number of photons passing through the cavity during one mechanical period is much larger than one. The quantum fluctuations in the photon number thus have a negligible effect on the dynamics of the condensate density modulations. In our simulations below, we have ensured that the bad cavity limit always applies.

In the first term inside the square bracket on the right-hand side of Eq.~(\ref{Eq:cavity}), $\Delta_c$ signifies the detuning of the driving field frequency from the cavity resonance frequency $\omega_{c}$.  The second term represents the coupling between the light mode and condensate, where the expectation value of the light potential $\cos^{2}\left(l\phi\right)$ taken with respect to the condensate wave function $\psi(\phi,\tau)$
\begin{equation}
    \langle \cos^{2}\left(l\phi\right)\rangle_{\tau}
    =\int_{0}^{2\pi}\left| \psi\left(\phi,\tau\right)\right|^{2}\cos^{2}\left(l\phi\right) d\phi,
\end{equation}
is a time-dependent quantity. In the third term, $\gamma_{0}$ is the energy decay rate of the cavity field. The last term inside the curly braces represents the laser drive with pump rate  $\eta = \sqrt{ P_{in} \gamma_{0} / \hbar \omega_c}$, where $P_{in}$ is the input optical power. The last term on the right-hand side of Eq.(\ref{Eq:cavity}) signifies the optical shot noise present in the system. The thermal and optical fluctuations each have zero mean and their correlations are given by~\cite{das2012winding,KumarPRL2021}
\begin{align} 
\langle\xi(\phi,\tau) \xi ^ *(\phi',\tau')\rangle  &= 2 \Gamma T \delta(\phi - \phi') \delta (\tau - \tau'), \\ 
\langle \alpha_{in}(\tau) \alpha_{in}^ *(\tau') \rangle  &= \omega_\beta\delta (\tau - \tau'), 
\end{align}
where $T$ is the non-dimensionalized temperature in units of $k_B/{(\hslash \omega_{\beta})}$, with $k_B$ being the Boltzmann constant. For the numerical simulation of these stochastic equations, the noise terms are modeled as follows
\begin{align} 
\xi(\phi,\tau)  &= \sqrt{2 \Gamma T / (d \phi d \tau)} \mathcal{N}(0,1,N_{\phi})  \mathcal{N}(0,1,N_{\phi}), \\ 
\alpha_{in}(\tau)  &= \sqrt{\omega _\beta / d\tau} \mathcal{N}(0,1,1), 
\end{align}
where $\mathcal{N}(0,1,N_{\phi})$ is a normally distributed random variable with zero mean and unit variance, where the third argument in $\mathcal{N}(0,1,N_{\phi})$ refers to the size of the array containing the random numbers and $N_{\phi}$ is the number of grid points in the $\phi$ direction. We have considered $N_{\phi}=1024$ for all the simulation runs.


To attain the dynamics of the persistent current, we have considered the initial state as a plane wave, and then we evolve the system in real-time using the coupled BEC - cavity equations and the RK4 scheme \cite{tan2012general}. A different approach is taken for the case of soliton. Initially, we prepare a localized state with a Gaussian density profile, and then we evolve it in imaginary time using the Strang splitting Fourier method \cite{bao2003numerical} to reach a soliton as its ground state in the presence of an optical lattice as well as atomic interactions. The resulting ground state is then used as the initial state for the subsequent real-time evolution using the RK4 scheme.  All the results in the paper have been presented for single realizations of a BEC, and each realization has been averaged over $5$ $\mu$s. We have adopted the time step $d\tau=10^{-7}$ for all the simulation runs.



\section{Results}

\subsection{Persistent current}
\label{PersistentCurrent}

\subsubsection{Rotational eigenstate}

In this section, we present the dynamics accompanying the detection of a persistent current in the ring BEC. Such currents can exist for macroscopic times as metastable flow states of the condensate with atoms that weakly repel each other \cite{RyuPRL2007}. The basic idea is for the circular optical lattice to act as a probe of the angular momentum, and hence the winding number, of the condensate \cite{KumarPRL2021}. For low intra-cavity photon number, the matter wave Bragg diffracts from the weak optical lattice. This results, in the first order, in two additional OAM states (sidemodes), which modulate the condensate density. These modulations add sidebands to the optical modes, which can subsequently be detected in the cavity transmission.

\begin{figure}[!htp]
\includegraphics[width=1\linewidth]{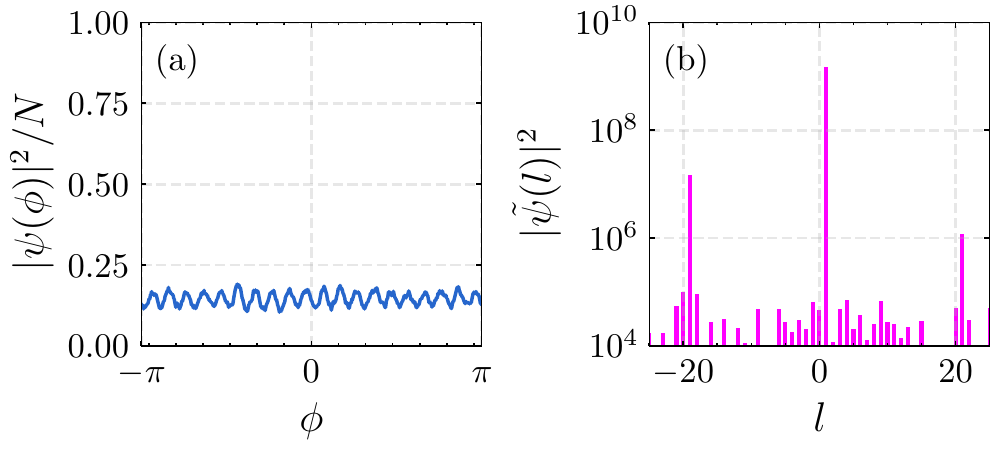} \\
\caption{(a) Angular profile of the condensate density per particle for a persistent current rotational eigenstate (b) OAM state content of the condensate. Parameters used here are $\Gamma = 0.0001$, $ T = 10$ nK  $ \times k_B/{(\hslash \omega_{\beta})}$ , $g / \hbar \simeq 2 \pi \times 0.02 $ Hz, $ L_p = 1$, $l = 10$, $N = 10^4$, $\Tilde{\Delta} = -2 \pi \times 173$ Hz, $ U_0 = 2\pi \times 212$ Hz, $ \gamma_{0} = 2\pi\times 2$ MHz, $P_{in} = 0.2$ pW, $\omega_c = 2\pi \times 10^{15}$ Hz, $m = 23$ amu, and $R = 12$ $\mu$m. }
  \label{fig:PW1}
\end{figure}

\begin{figure}[!htp]
\centering
\includegraphics[width= 1\linewidth]{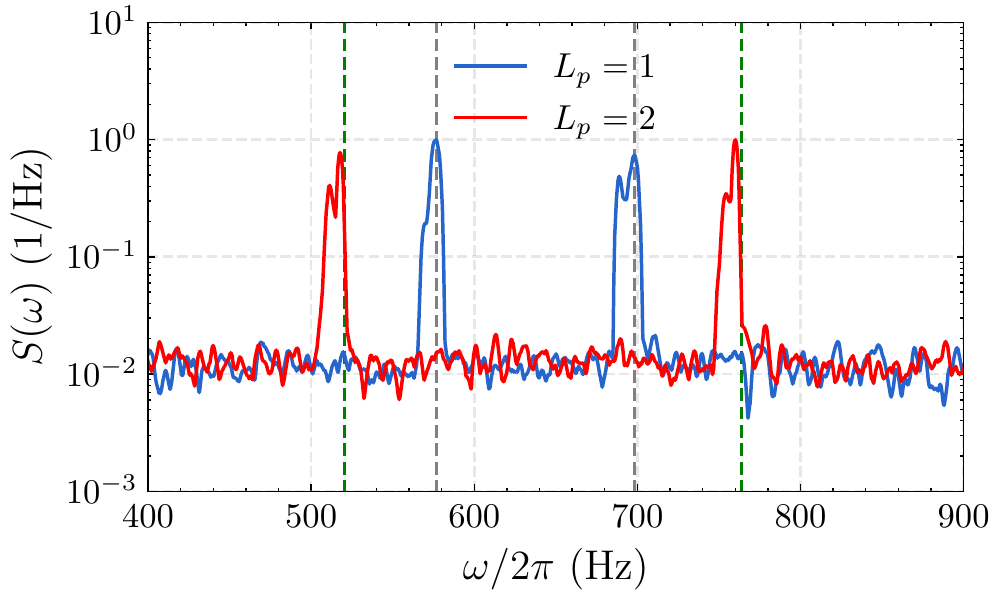} \\
\caption{Power spectra of the output phase quadrature of the cavity field as a function of the system response frequency for $L_{p}=1$ (blue) and 
$L_{p}=2$ (red). The vertical dashed lines (grey and green) correspond to the analytical predictions for the side modes of $L_p = 1$ and $L_p = 2$ respectively, including atomic interactions, made earlier \cite{KumarPRL2021}. 
Other parameters are same as mentioned in Fig. \ref{fig:PW1}.}
\label{fig:PW2}
\end{figure}


\begin{figure}[!htp]
\centering
\includegraphics[width= 1 \linewidth]{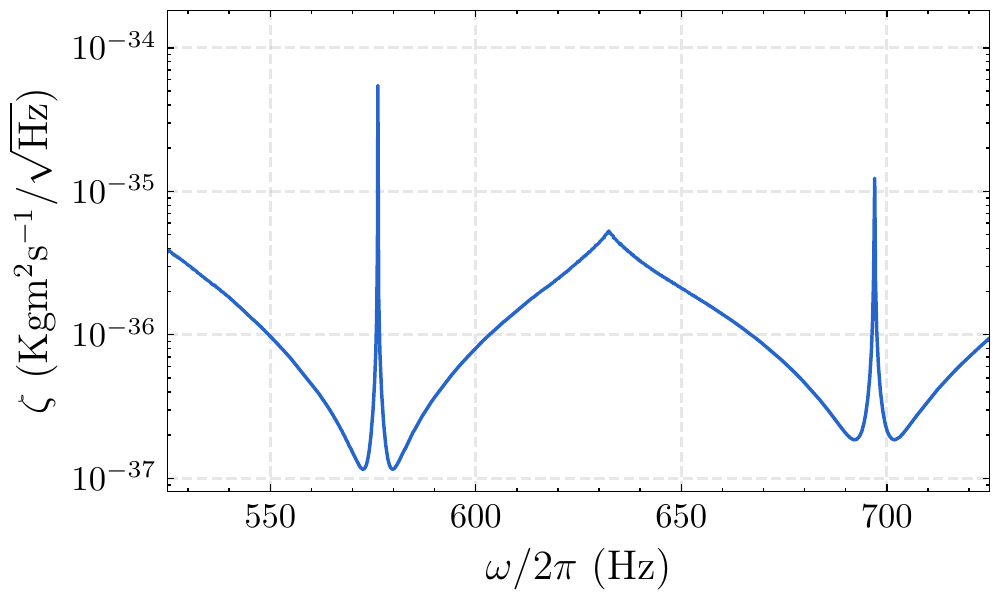} \\

\caption{Persistent current eigenstate rotation measurement sensitivity $\zeta$ [Eq.~(\ref{eq:Sensitivity})] as a function of the system response frequency $\omega$. Here $G = 2 \pi \times 7.5$ kHz and $|\alpha_s|^2 = 0.096$, which corresponds to $P_{in} = 0.2$ pW. Other parameters are the same as for Fig. \ref{fig:PW1}.}
  \label{fig:PW4}
\end{figure}

In our simulation, for which we have used $^{23}$Na atoms \cite{RyuPRL2007}, a phase gradient is imprinted initially on the condensate, in order to impart a winding number $L_{p}$ to it. The resulting persistent current then gets coupled to the angular optical lattice, which displays $2l$ interference maxima along the ring on which the BEC is trapped. For all our simulations related to the persistent current, we consider an initial state for the condensate wave function of the form
\begin{equation}
    \psi(\phi) = \sqrt{\frac{N}{2 \pi}} e^{i L_p \phi},
\end{equation}
which corresponds to an eigenstate of condensate rotation in the absence of the optical lattice.

\begin{figure}[!htp]
\centering
\includegraphics[width= 1 \linewidth]{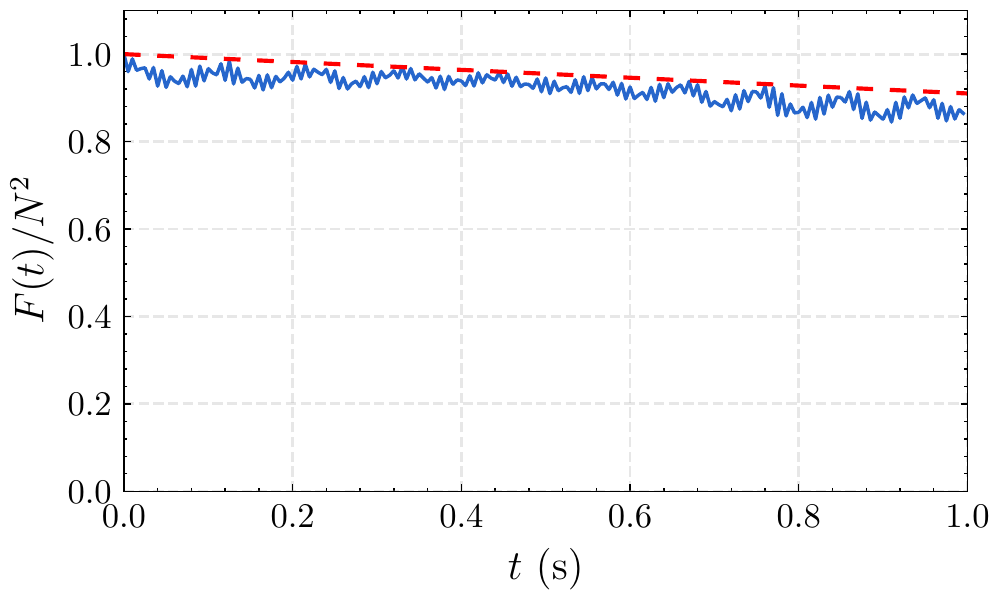} \\
\caption{Variation of the fidelity [Eq.~(\ref{eq:Fidelity})] of the ring condensate density for a persistent current with time. The red dashed line is the guide to see the fidelity value when the density profile of the condensate at different times are in phase with the initial state. The set of parameters are same as those in Fig.~\ref{fig:PW1}.}
\label{fig:PW5}
\end{figure}

The resulting condensate density ($\psi$) obtained from the Eqs.~(\ref{Eq:bec})-(\ref{Eq:cavity}), modulated by the presence of the condensate sidemodes created by the optical lattice, is shown in Fig.~\ref{fig:PW1}(a). The OAM content of the modulated condensate density ($\lvert\tilde{\psi}\rvert^2$), where $\tilde{\psi}$ is the Fourier amplitude of the condensate density ($\psi$), is shown in Fig.~\ref{fig:PW1}(b), which displays the first-order peaks, resulting from matter-wave Bragg diffraction, at $L_{p}\pm 2l$. The figure, which accounts for the full Gross-Pitaevskii condensate dynamics, implies that only three OAM modes are dominant and therefore provides justification for the few-mode model proposed earlier. \cite{KumarPRL2021}. 


\begin{figure}[b]

\includegraphics[width= 1 \linewidth]{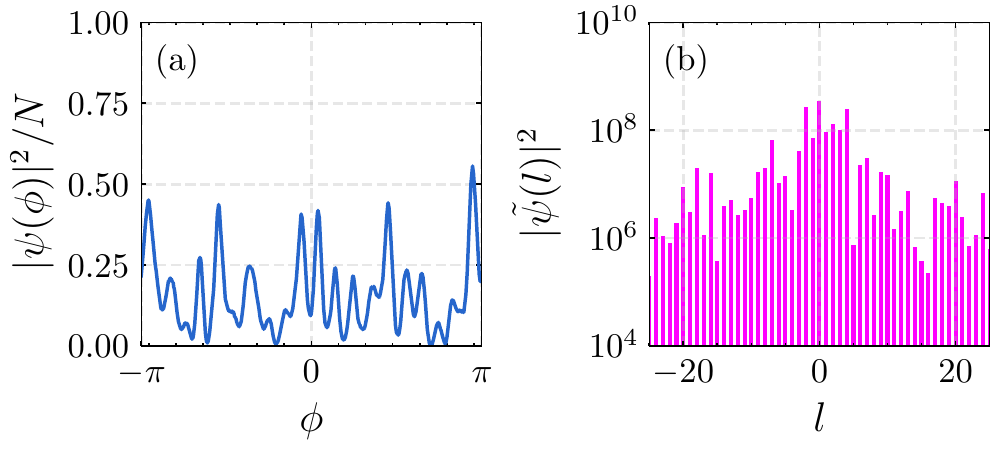} \\
\caption{(a) Angular profile of the condensate density per particle for a persistent current superposition [Eq.~(\ref{eq:Super})] with $L_{p1} = 3, L_{p2} = 4$ (b) OAM state content of the condensate. Here $P_{in} = 1$ pW and the other parameters are the same as in Fig.~\ref{fig:PW1}.}
\label{fig:PW6}
\end{figure}

In Fig.~\ref{fig:PW2} we show the phase quadrature of the resulting cavity transmission spectrum \cite{KumarPRL2021},
\begin{equation}
\label{eq:Spect}
S(\omega)=\left|\mathrm{Im}\left[\alpha_{out}(\omega)\right]\right|^{2},
\end{equation}
where $\alpha_{out}$ is the output field, transmitted from the cavity and it is related to the input field into the cavity, through the input-output relation $\alpha_{out} = -\alpha_{in} + \sqrt{\gamma_0}\alpha$ \cite{AspelmeyerRMP2014}. The vertical dashed lines for $L_{p}=1$ correspond to the sidemode frequencies $\omega_{c,d}$ defined in the Supplementary Material of \cite{KumarPRL2021}, see Eq.S33. These definitions account for the atomic interactions. The agreement between the vertical lines and the peak locations show that the Gross-Pitaevskii simulation retains the results predicted by the few-mode theory presented earlier. It also shows that our classical treatment of $\alpha$, the optical field, reproduces the results of \cite{KumarPRL2021}, which treated the optical field quantum mechanically. It can also be seen that the peaks for $L_{p}=2$ are spectrally distinct from those for $L_{p}=1$. Thus our method can reliably distinguish between neighboring values of condensate winding number $L_{p}$. We have also analyzed the effect of high cavity power drive $P_{in}$ on the power spectra of the output phase of the cavity and find that increasing the $P_{in}$ leads to the deviation of the different peaks of $S(\omega)$ from the analytical results as well as generation of the other OAM modes at higher frequencies as shown in the Fig.~\ref{fig:sol4}.

\begin{figure}[!htp]

\includegraphics[width= 1 \linewidth]{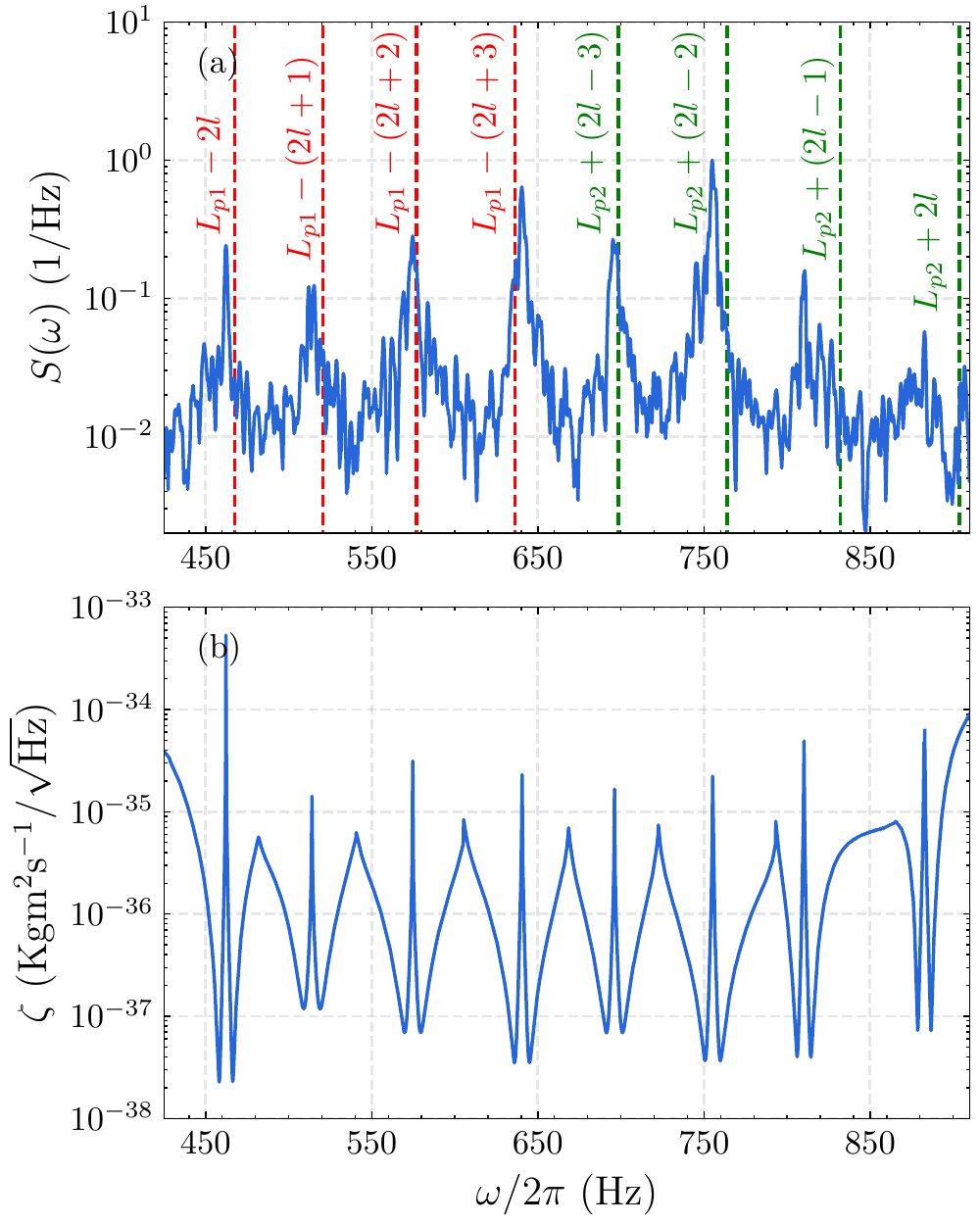} \\

\caption{Persistent current superposition (a) Power spectrum of the output phase quadrature of the cavity field as a function of the system response frequency. The vertical dashed lines correspond to the analytical predictions for the side modes corresponding to $L_{p1} = 3, L_{p2} = 4$.   (b) Rotation measurement sensitivity. The parameters used are the same as in Fig.~\ref{fig:PW6}.}
\label{fig:PW7}
\end{figure} 

\begin{figure}[!htp]

\includegraphics[width= 1.0 \linewidth]{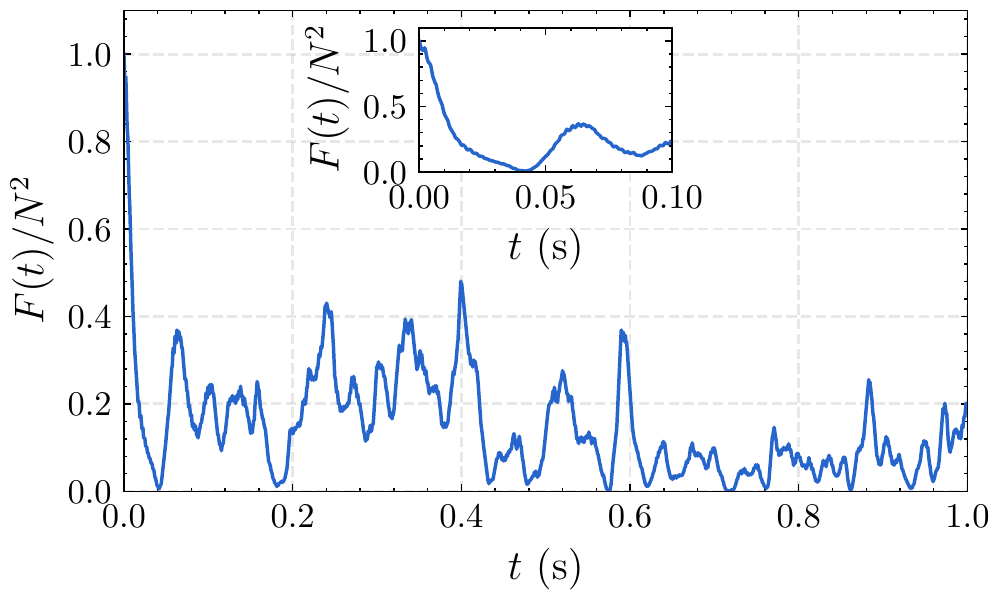} \\
\caption{Variation of the fidelity [Eq.~(\ref{eq:Fidelity})] of the condensate density with time, resulting from the superposition of two persistent currents. The set of parameters are same as in Fig.~\ref{fig:PW6}. Inset shows the fidelity from the initial time to
$100$ ms.}
\label{fig:PW8}
\end{figure}

We quantify the performance of our scheme by using the sensitivity of the rotation measurement, defined as 
\begin{equation}
\label{eq:Sensitivity}
    \zeta = \frac{S(\omega)}{\partial S(\omega)/\partial\Lambda} \times \sqrt{t_{meas}},
\end{equation}
where $t_{meas}^{-1} = 8 (\alpha_s G)^2 / \gamma_0$ is the optomechanical measurement rate in the bad cavity limit, $ G = U_0 \sqrt{N} / 2 \sqrt{2}$ \cite{KumarPRL2021}, and $\alpha_s$ is the steady state of the cavity field \cite{KumarPRL2021}. The sensitivity of Eq.~(\ref{eq:Sensitivity}), obtained from a fit to the spectrum [Eq.~(\ref{eq:Spect})], is displayed in Fig.~\ref{fig:PW4}, and matches the result from \cite{KumarPRL2021} quite well.

To quantify to what extent the measurement backaction affects the condensate, we display the fidelity $F(t)$, i.e. the position-averaged autocorrelation function of the condensate density
\begin{equation}
\label{eq:Fidelity}
    F(t) = \int_{0}^{2\pi} \left[\psi^{*}(\phi,t)\psi(\phi,0)\right]^2 d\phi, 
\end{equation}
as a function of the unscaled time $t$ in Fig.~\ref{fig:PW5}. As can be seen, the density fidelity stays close to unity for small times. For macroscopic times, it decays mainly due to the damping ($\Gamma$) and noise
($\xi$) of the persistent current. This indicates that the measurement backaction on the condensate is small. Certainly, unlike existing techniques, the measurement does not completely destroy the condensate.

\subsubsection{Two state superposition}

In this section, we investigate the dynamics of the condensate prepared in a superposition state of two different winding numbers i.e. $L_{p1} \neq L_{p2}$. These states could be of interest in the context of quantum information processing, matter wave interferometry, as well as studies of mesoscopic quantum mechanics and \cite{HallwoodPRA2010,AndersenPRL2006}. We start with the initial state
\begin{equation}
\label{eq:Super}
   \psi(\phi) = \sqrt{\frac{N}{4 \pi}} 
   \left(e^{i L_{p1}\phi}+e^{iL_{p2}\phi}\right).
\end{equation}

Fig. \ref{fig:PW6}(a) shows the condensate density resulting from the superposition of two persistent currents. The modulation in the condensate density is relatively high due to the high input optical power required to observe condensate rotation. The OAM content of the corresponding state is shown in Fig. \ref{fig:PW6}(b). The increased complexity in the OAM content of the state is due to the interference between the two persistent currents, which introduces additional modulations to the system. Despite these complicated modulations, sidebands corresponding to $ L_{p1}$ and $L_{p2}$ can be observed in the phase quadrature of the resulting cavity transmission spectrum, which is presented in Fig. \ref{fig:PW7}(a). 

The sensitivity of the rotation measurement is shown in Fig. \ref{fig:PW7}(b) and is comparable to that of the measurement of the rotational eigenstate [Fig.~\ref{fig:PW4}]. Variation of the fidelity of the condensate density with time is shown in Fig. \ref{fig:PW8}. In this case, due to the interference between multiple sidemodes, the fidelity of the superposition measurement is at best around $0.5$.

\subsection{Bright soliton}
\label{Soliton}
In this section, we present the dynamics accompanying the detection of a bright soliton in the ring BEC. Such solitons can be sustained by the condensates, in which the atoms weakly attract each other~\cite{PethickBECBook,RuprechtPRA1995,GarciaPRA1998,KhaykovichScience2002,StreckerNature2002,CornishPRL2006,MarchantNC2013,MedleyPRL2014,LepoutrePRA2016,MeznarsicPRA2019}. In our simulations, we imprint a density and phase modulation on a uniform condensate of $^{7}$Li atoms \cite{MedleyPRL2014} and this leads to a bright soliton rotating on the ring and carrying a winding number $L_{p}$ (e.g. see Eq.~(43) of \cite{KanamotoPRA2003}). For this situation, we consider the initial state as
\begin{equation}
    \psi(\phi) = \sqrt{\frac{N}{\sqrt{\pi}}} e^{- \phi^2 / 2} e^{i L_p \phi}.
\end{equation}

\begin{figure}[!htbp]
\includegraphics[width= 1.0 \linewidth]{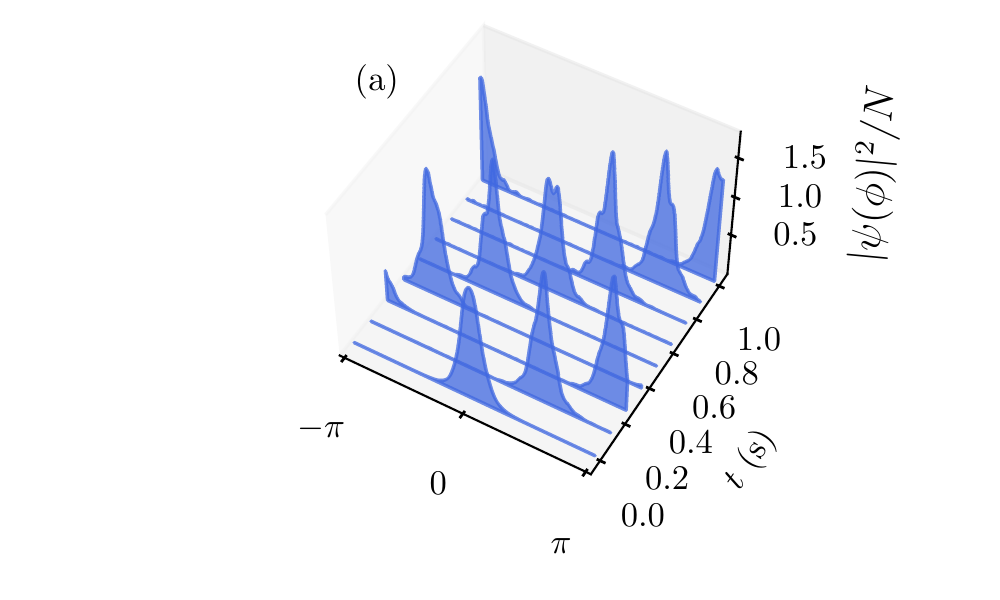} \\
\includegraphics[width= 1 \linewidth]{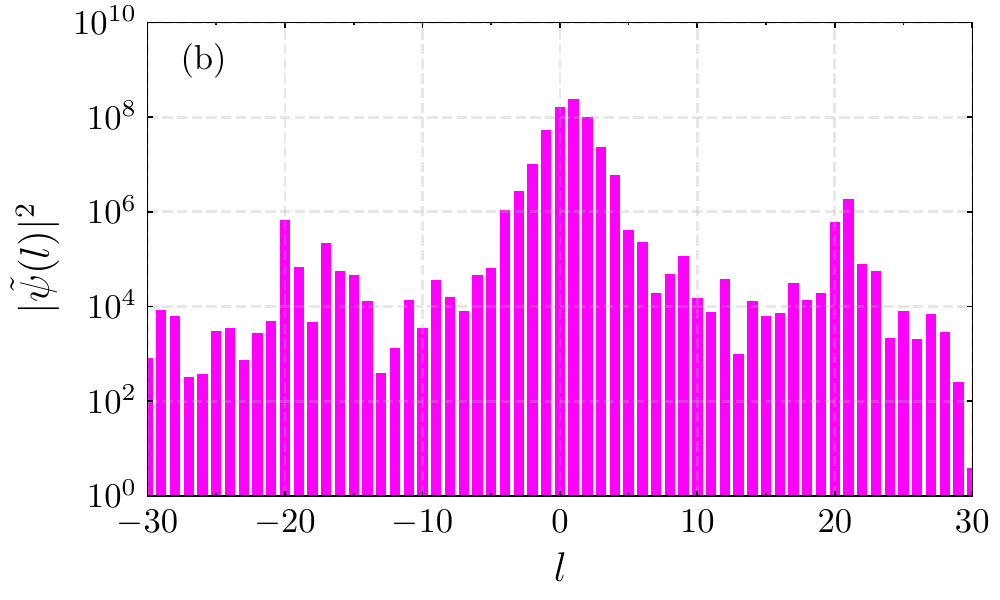} \\
\caption{(a) Temporal evolution of the moving soliton density profile (b) OAM content of the soliton. Here $N = 6000$, $a_s = -27.6 a_0 $, where $a_0$ is the Bohr radius, $m = 7.01$ amu, $L_p = 1$,  and $P_{in} = 0.4$ pW, and all other parameters are same as in Fig. \ref{fig:PW1}.}
\label{fig:sol2}
\end{figure}

\begin{figure}[!htp]

\includegraphics[width=\linewidth]{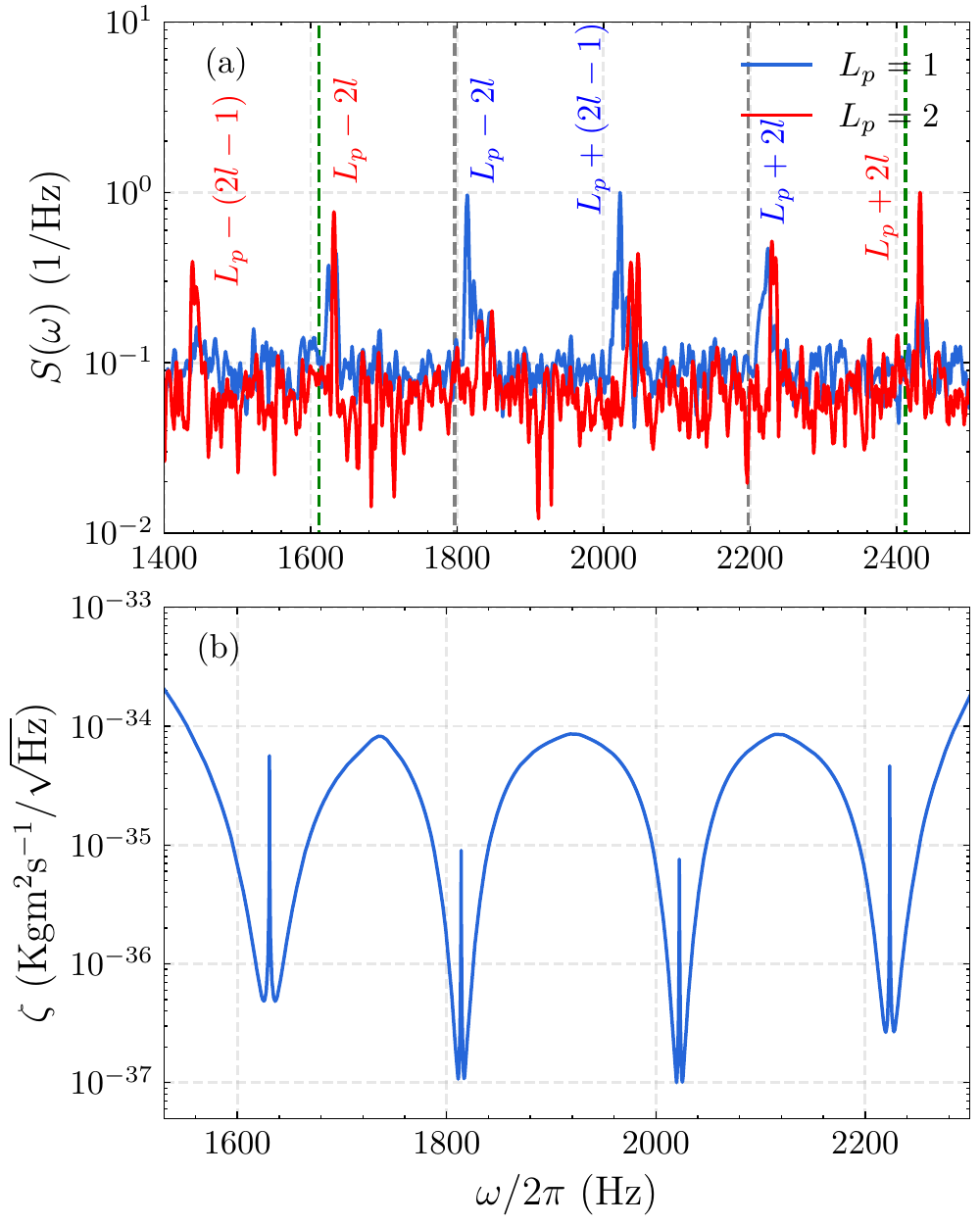} \\

\caption{(a) Power spectrum of the output phase quadrature of the cavity field as a function of the system response frequency for a soliton. The vertical dashed lines (grey and green) correspond to the analytical predictions for the side modes of $L_p = 1$ and $L_p = 2$ respectively. (b) Variation of rotation measurement sensitivity  $\zeta$ with the system response frequency $\omega$ for $L_p = 1$.  The set of parameters used here are the same as in Fig. \ref{fig:sol2}.}
  \label{fig:sol3}
\end{figure}

For $L_{p}=1$, the temporal evolution of the soliton density profile is shown in Fig.~\ref{fig:sol2}(a). As can be seen, the spatial profile of the soliton stays close to its initial shape as it moves in the ring. The OAM distribution of the soliton, when it has interacted with the optical lattice, is shown in Fig.~\ref{fig:sol2}(b).

The resulting cavity transmission spectrum  used to detect the winding number of the soliton is shown in Fig.~\ref{fig:sol3}(a). As can be seen, the $L_{p}=1$ and $L_{p}=2$ peaks are resolvable, indicating that our method can distinguish between neighboring winding numbers for the soliton. Remarkably, the analytical predictions for the spectral locations of the sidemode peaks from the analytical treatment of the persistent current case  \cite{KumarPRL2021} agree quite well with the full Gross-Pitaevskii treatment of the soliton. This can be seen from the coincidence of the numerically obtained side mode peaks and the vertical dashed lines in Fig.~\ref{fig:sol3}(a).

\begin{figure}[!htbp]

\includegraphics[width= 1.0\linewidth]{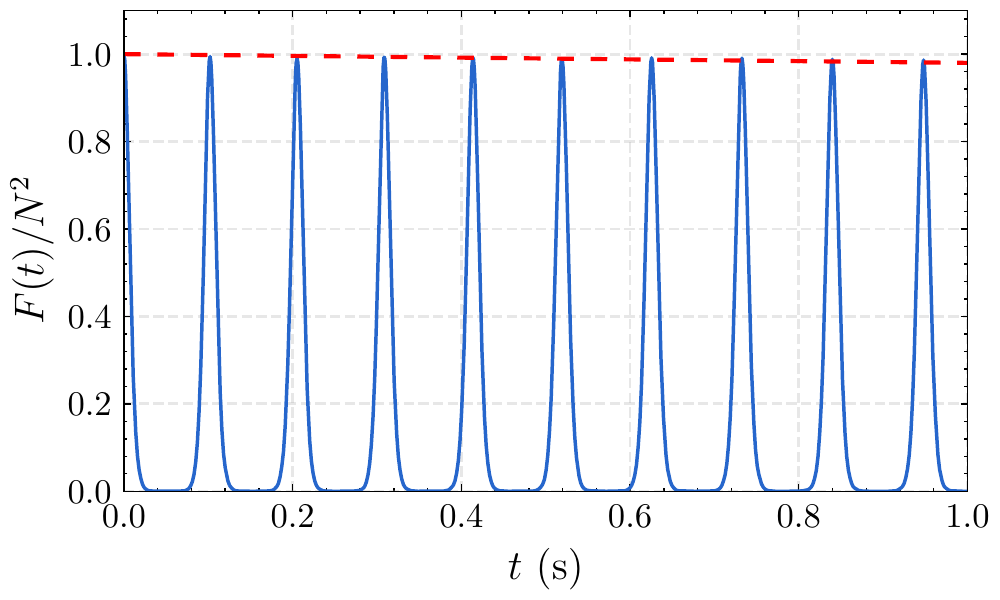} \\

\caption{Fidelity (Eq.~\ref{eq:Fidelity}) of the soliton density profile versus time for $P_{in} = 0.4$ pW. The red dashed line is the guide to the fidelity value when the solitons at different times are in phase with the initial state. The remaining set of parameters are the same as in Fig. \ref{fig:sol2}.}
  \label{fig:sol5}
\end{figure}

The corresponding measurement sensitivity is shown in Fig.~\ref{fig:sol3}(b). As can be seen, the  benefits of high measurement sensitivities in our optomechanical scheme carry over from the persistent current case (where it was three orders of magnitude better than demonstrated previously \cite{KumarPRL2021}) to the case of the bright soliton. The fidelity of the density profile is shown in Fig.~\ref{fig:sol5} and remains close to unity for macroscopic times. The decay in fidelity is largely due to the dissipation ($\Gamma$) and noise ($\xi$) in the system, and not so much due to measurement backaction. Hence our scheme represents a minimally destructive measurement of the motion of a bright soliton in a ring BEC. For high input powers results the appearance of more modes accompanied with the noise in the condensate as shown in the Fig.~\ref{fig:sol4}.

\section{Summary and Conclusions}
\label{sec:5}
Using a stochastic mean-field Gross-Pitaevskii formalism for modeling atoms in a BEC in a ring trap, and a classical approximation for the optical mode, we have demonstrated that cavity optomechanics can make real-time, \textit{in situ}, and minimally destructive measurements of both persistent currents as well as bright solitons. In support of our conclusions, we have presented numerical simulations of cavity transmission spectra, measurement sensitivities as a function of the system response frequency, and the fidelity of condensate density profiles. 

Our numerical simulations have verified and extended the analytical model proposed by us earlier. Remarkably, the previously analytically-found locations of the peaks in the cavity transmission crucial for determining condensate rotation agree well with the numerical results for both persistent current states as well as solitons. 

We expect our findings to be of interest to studies of superfluid hydrodynamics, atomtronics, and soliton interferometry. The technique we have presented could be extended to other systems such as polariton ring condensates \cite{SaitoPRB2016}.
\section{Acknowledgments}
We thank the International Centre for Theoretical Sciences, Bengaluru, where this work was initiated, for hosting us. M.B. would like to thank the Air Force Office of Scientific Research (FA9550-23-1-0259) for support. R.K. acknowledges support from JSPS KAKENHI Grant No. JP21K03421. We also gratefully acknowledge our supercomputing facility Param-Ishan (IITG), where all the simulation runs were performed.
\appendix
\counterwithin{figure}{section}
\section{Noise for high input power}
\label{HighSpec}
In this appendix, we present the effect of increasing the cavity drive power $P_{in}$ and thus making a stronger measurement of the persistent current and bright soliton rotation. 

\begin{figure}[!htbp]
\centering
\includegraphics[width= 1.0 \linewidth]{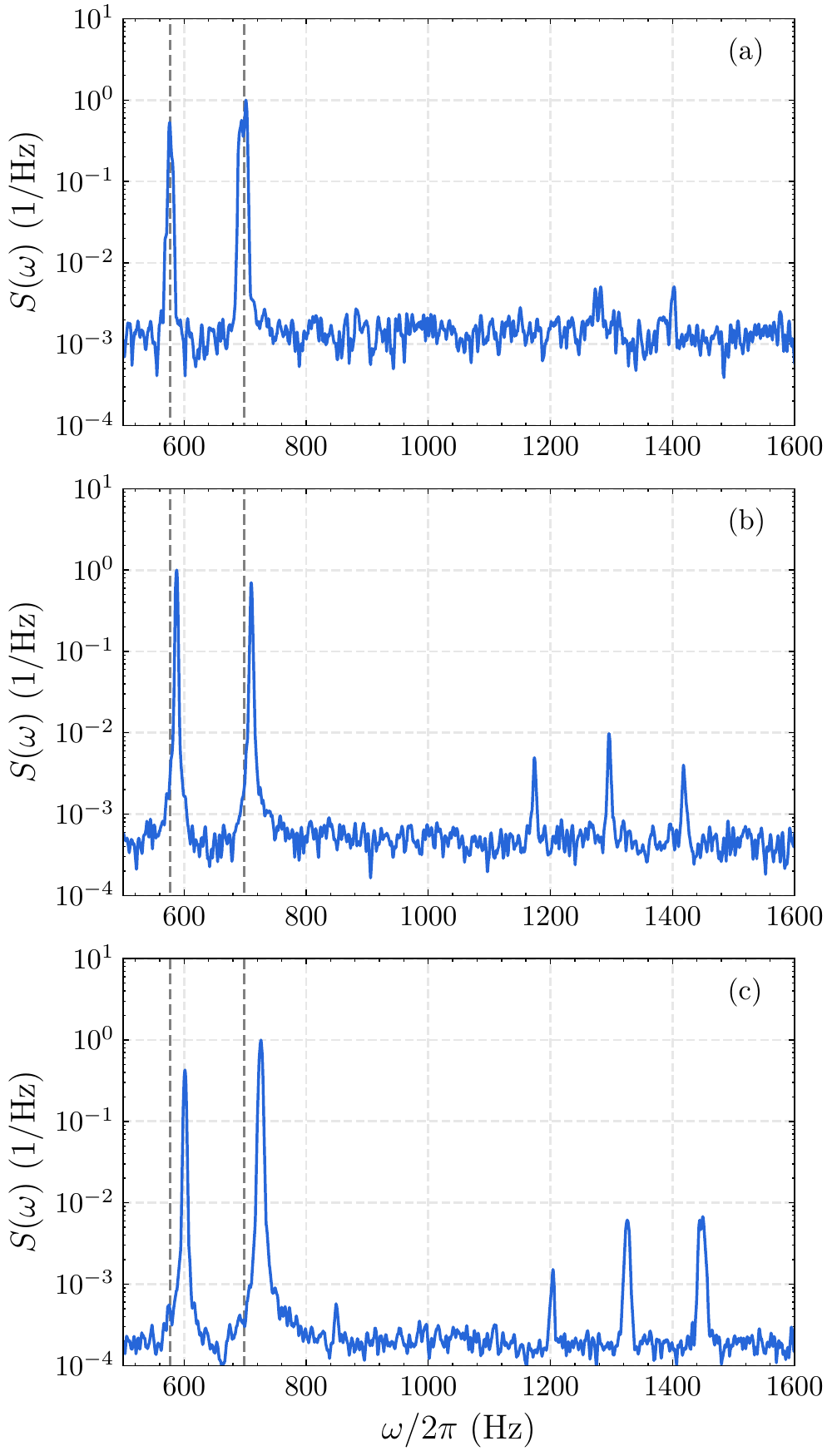} \\

\caption{Power spectrum of the output phase quadrature of the cavity field as a function of the system response frequency for different input powers, 
(a) $P_{in} =0.5$ pW, (b) $P_{in} =1$ pW, and (c) $P_{in} =2$ pW. Vertical dashed lines indicate the analytical predictions of \cite{KumarPRL2021}. All other parameters are the same as in Fig. \ref{fig:PW1}.}
  \label{fig:PW3}
\end{figure}

In  Fig.~\ref{fig:PW3}, we show the power spectra of the output phase quadrature of the cavity field in the frequency domain for three different input powers, namely, $P_{in}=0.5$ pW, $1$ pW and $2$ pW for the persistent current. We find that increasing the power (from the left column to the right) results in nonlinear behavior, as discussed below.

\begin{figure*}[!htb]
\centering
\includegraphics[width=1\linewidth]{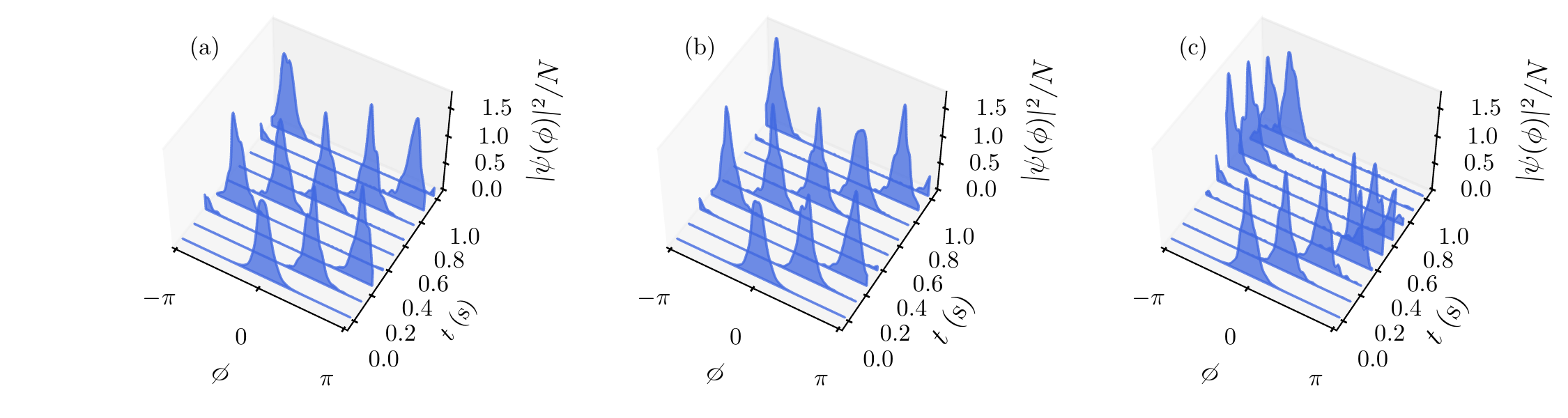} 
\includegraphics[width=1\linewidth]{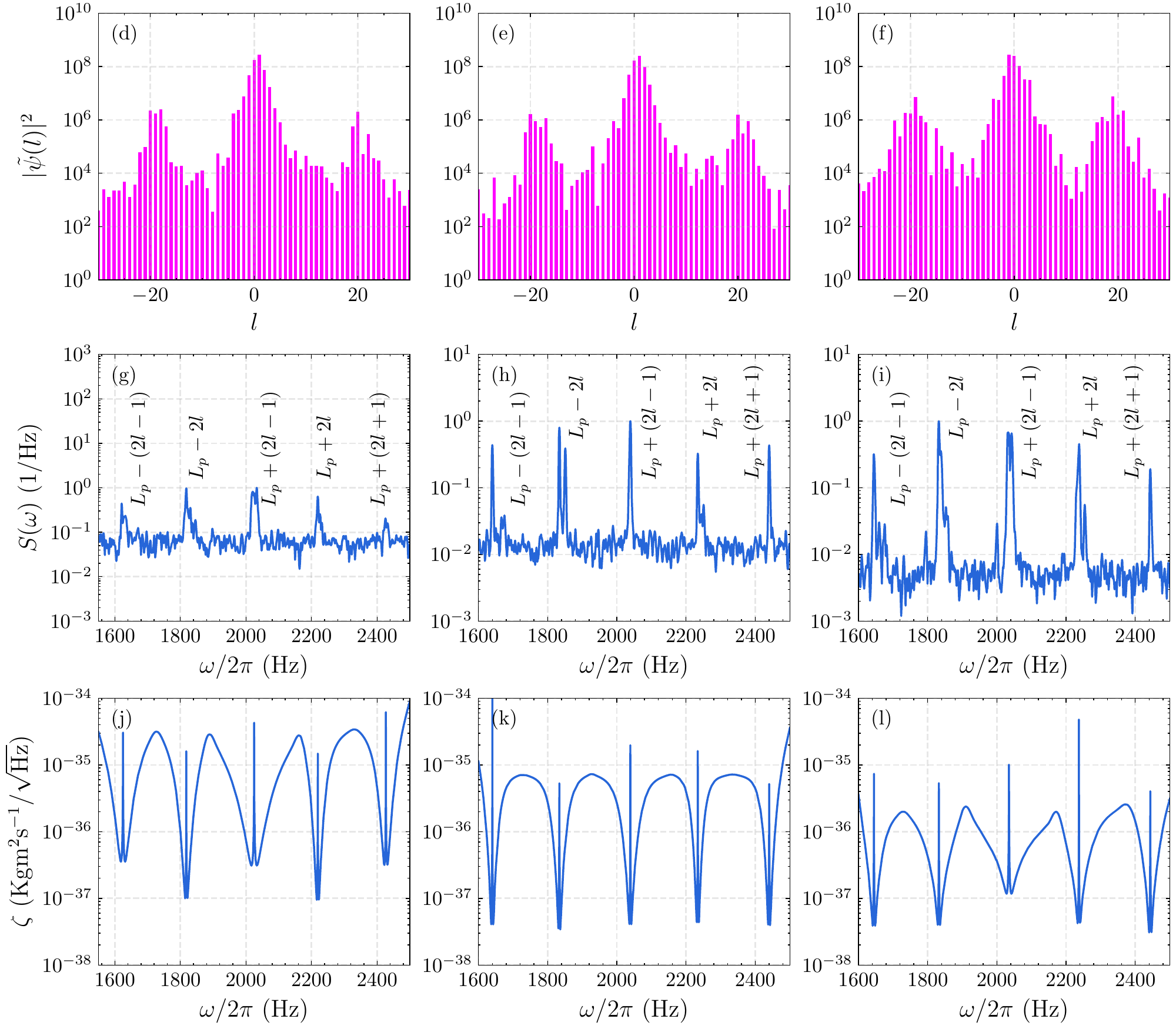} \\

\caption{ (a)-(c)Temporal evolution of the moving soliton density profile (d)-(f) OAM states of the condensate occupied by the soliton (g)-(i) the power spectrum of the imaginary part of cavity field versus response frequency (j)-(l) soliton rotation measurement sensitivity as a function of system response frequency for $P_{in} = 0.7$ pW, $1$ pW, and $2$ pW, respectively. Here $G = 2 \pi \times 5.8$ kHz and $|\alpha_s|^2 = 0.33, 0.48, 0.96$ for the above input power values. The remaining set of parameters are the same as in Fig. \ref{fig:sol2}.}
    \label{fig:sol4}
\end{figure*}

Density profiles of the soliton have been shown in Figs.~\ref{fig:sol4}(a)-(c). We can see that at low powers $P_{in}$ the density profile is only slightly modulated, while it is quite heavily modulated at high $P_{in}$. Thus, as expected, the measurement backaction increases with cavity power. However, in the regime of the powers presented, the soliton does not break up as a result of interaction with the probe lattice, and thus the measurement is not fully destructive. 

The OAM content of the soliton for the corresponding powers have been shown in Figs.~\ref{fig:sol4}(d)-(f). The spectra are displayed in Figs.~\ref{fig:sol4}(g)-(i), the peaks labeled by the winding number $L_{p}$. As can be seen, use of higher optical powers results in additional peaks and more noise. Fianlly the sensitivities are shown in Figs.~\ref{fig:sol4}(j)-(l), showing that they are comparable
to the low power version.

\clearpage
\bibliography{citation.bib} 



\end{document}